\documentclass[9pt,twocolumn,twoside]{pnas-new}
\usepackage{graphics}
\usepackage[outdir=./]{epstopdf}
\templatetype{pnasresearcharticle}  
\setboolean{displaywatermark}{false}
\title{coupling between risk and cautious behavior affect epidemic morbidity, mortality and dynamics.}
\author[a]{Yoel Sanders}

\affil[a]{Department of Mathematics, Bar-Ilan University,
Ramat-Gan, 52900, Israel}

\leadauthor{Sanders} 
\significancestatement{Models for epidemic spread typically account for variable risk factors but do not account for the correlation between behavior and risk. Here we extend these models to account for such correlations. We find that a positive correlation between behavior and risk, i.e., voluntary risk aversion by individuals at high risk and risk-taking behavior by individuals at low risk, leads to a linear reduction in morbidity and mortality and load on healthcare services compared to the uncorrelated case. We show that increasing caution in response to news from countries with a preceding outbreak leads to a more graded response to a lock-down. We also show that if vaccinated individuals are less cautious an increase in herd immunity threshold ensues. 
}

\authorcontributions{}
\authordeclaration{Authors declare no conflict of interest.}
\correspondingauthor{\textsuperscript{1}To whom correspondence should be addressed. E-mail: yoelsanders@gmail.com}


\begin{abstract}
Models for epidemic spread evaluate the risk of capacity-overflow of public healthcare services and facilitate decisions regarding preferred mitigation strategies. Current epidemic spread models take variable risk factors such as age or medical preconditions into account, but they typically do not account for the coupling between behavior and risk. Behavior and risk are coupled, e.g., by individuals voluntarily wearing masks or reducing social contacts due to a perceived high risk or by individuals who avoid these actions due to perceived low risk. Although game-theoretic and economic models account for such coupling between behavior and risk, they require a utility function as input. However, the utility function is typically unknown and hard to estimate. Here we offer a simple model that quantitatively accounts for the coupling between risk and behavior in a manner that can be readily estimated. We demonstrate that such a coupling can have significant consequences on the course of an epidemic. Positive coupling between risk and behavior can lead to a significant reduction in healthcare load and to mortality compared to the uncoupled case. Dynamic change in the risk aversion behavior in response to a change in perceived risk can lead to a graded and delayed response in the number of hospitalized patients after the enforcement of a lockdown. Finally we show that a reduction in caution after vaccination can lead to an increase in the herd immunity threshold. \end{abstract}

\dates{This manuscript was compiled on \today}

\begin{document}

\maketitle
\thispagestyle{firststyle}
\ifthenelse{\boolean{shortarticle}}{\ifthenelse{\boolean{singlecolumn}}{\abscontentformatted}{\abscontent}}{}

\section *{introduction}
\dropcap{T}he simplest model for epidemic spread 
consider a population that is coarsely subdivided  to three mutually exclusive groups; susceptible individuals that did not contracted the disease (S), infected individuals that are currently infectious (I), and recovered individuals that are can no longer infect others (R). 

A well-known model of this sort it the SIR model which further assumes mass-action kinetics i.e., well mixing between all sub-populations, and so the rate of depletion of susceptible individuals is proportional to the multiplication of the number of infected individuals with the number of susceptible individuals, $I \times S$, with a proportionality rate constant $\beta$ that equals to the rate of infection per individual in society. The second parameter that enters the model is $\gamma^{-1}_{R}$---the average duration of being infectious. The average number an infected individuals infects---the basic reproductive number $R_0$---is the infectious rate per individual times the average infectious period, i.e., $R_0=\beta \gamma_R^{-1}$ \cite{R0Definition}. If the distribution of the number of people infected by an infected individual is broad, i.e., has large coefficient of variation, or fat-tailed, e.g., has a diverging mean, or if the distribution of the number of people is non-stationary, then $R_0$ is no longer a good summary of the situation. In these circumstances the dynamics can become non-ergodic, and use of other parameters that characterize the distribution becomes essential \cite{CovidInfectionOverTime}, \cite{CM}, \cite{superspreadingVariance} [ref:muli12, taleb, goldenfeld12]. 

The SIR model is the starting point for more elaborate set of models, known collectively as compartmental models \cite{CompartmentModelSummary}, which are characterized by additional states or compartments. States are added to refine the level of description. For example, to facilitate a distinction between recovery and death, to distinguish between symptomatic and non-symptomatic infections or to add an incubation period. In many diseases, the infectivity profile is not constant in time and follows a certain profile \cite{XE}. To better account for an inhomogeneous temporal infectivity profile, states can be added to describe the progression of infectivity during the infectious period \cite{covidInfectionProfileCorrection}.  

Compartmental models can be viewed as a mean-field limit of stochastic epidemiological models defined on social structured connectivity graphs \cite{meanFieldConvergenceCovid}. Previous works have studied the interplay between the mean-field parameters of the SIR model, especially the infection rate per individual $\beta$ and graph parameters such as its connectivity distribution \cite{PDEvsRandomGraphs} \cite{networksVsoutbreaks} or its clustering coefficient \cite{CM}. Mean-field models are not suitable for studying small-number effects such as the prediction of epidemic onset, or when the distribution of contacts has a large coefficient of variation or worse---infinite variance or infinite mean and variance [ref123]. 

Adding more states to a compartmental model is a mixed blessing. Inevitably, adding more states requires adding more rate parameters. To extract predictions from the model, these rate parameters requires estimation. For example, differentiating between symptomatic and non-symptomatic patients requires adding different rate constants that describe the variation in infectivity, hospitalization, recovery, and death between the two groups. Furthermore, adding more states require estimating their value at a given time point in order to form the estimated initial condition. However, addition of more states can enable new understanding of mechanisms and relations that are otherwise hidden in coarser models. Finding the right balance between oversimplification to excessive modeling remains a major challenge for analysts. 

The initial dynamics of an epidemic outbreak is exponential growth in the number of infected individuals \cite{exponentialGrowthEpidemicOutbreak}. Exponential growth dynamics renders the prediction of the turning point---the point in time where the percent of the population infected reaches its peak---inherently difficult as it is highly sensitive to the unknown data and model parameters \cite{SMM}. For example, the total number of infected individuals at the onset of an outbreak, is typically unknown and the uncertainty can be very high (i.e., above $100\%$). Similarly, the uncertainty in infectivity at the onset of an outbreak is also high, especially when new strains are involved. The exponential sensitivity of the predicted turning point to these unknown data and parameters are bad news for analysts and decision-makers alike \cite{epidemicThresholdSensitivityPNAS}. The sensitivity can render analysts initially unable to improve the accuracy or uncertainty of their predictions. 

In this work, we present a simple extension relevant to numerous compartmental models, that demonstrate how the coupling between elective cautious behaviour of individuals in society and their actual risk, affect peak hospital load, overall mortality, dynamics after onset of a lockdown, and modulation of the herd immunity threshold during a vaccination campaign. 

Risk aversion or lack of risk aversion by individuals in society is related to their perceived risk. Individuals at risk can either recognize their risk and as a result use masks in public or reduce social contacts, or fail to do and hence raise their chance of infection. To demonstrate the effect of such dichotomy we present a modified SIR model with four quadrants; (i) non cautious individuals at risk, (ii) non-cautious individuals not at risk, (iii) cautious individuals not at risk, and (iv) cautious individuals at risk. We denote the size of group (i)---non-cautious individuals at risk---as \(\Phi\). This group, causes the largest per-capita stress on the healthcare infrastructure \cite{JRG}.

In the current SARS-Cov-2 epidemic, risk factors (age in particular), strongly affect the hospitalization and mortality rates. \cite{JZ} proposes a method for estimating the age-to-age infection rate constants based on extensive surveys. In our model, we coarse-grained the population into two groups with respect to risk, high and low risk which roughly corresponds to above the age of $60$, and below the age of $60$ respectively. Similarly, we coarse-grained the population into two groups, cautious, and not-cautious. More refined distinction is of course possible. Our goal is to demonstrate the effect of coupling between risk and behaviour of the severity and dynamics of an epidemic, and for such a purpose, we find that binary distinction is sufficient.

In the context of the current epidemic, the dominant risk factor for Covid-19 related hospital admission is age and pre-existing medical condition, and the most simple risk aversion strategy to be taken is public mask wearing. According to CDC \cite{particleSpreadCovid} wearing a mask reduces the risk of infection by $\sim 75\%$ in a risky encounter with a infectious individual that does not wear a mask, and by $\sim 85\%$ if mask wearing is mutual. We consider mask wearing cautious behaviour and split the population into cautious and not cautious. 


Here we argue that the an overlap between the population fraction at risk with the population that exhibit cautious behaviour, affects the severity, in terms of morbidity and mortality, as well as the dynamics of the epidemic. Societies where the perceived risk by members is correlated with their actual risk results in less load on hospitals and less mortality compared to the uncorrelated or anti-correlated case. We explain how estimation of the level of coupling between risk aversion behaviour and actual risk in the current pandemic, is possible via analyzing vaccination records per age groups in the population.

\section*{Results}
We find that positive coupling between risk and behaviour can lead to a reduction in the overall load to hospitals. In Covid-19, individuals at risk have at least an order of magnitude more chance of being admitted to an hospital. Thus, the benefit of mitigating risk by the sub-population at risk is larger than there share in the population \cite{XE}. 

To understand the effect of coupling between behaviour and risk on hospital load and mortality we used infectious rate parameter and infectious time window that are similar to the current Covid-19 pandemic, and scanned the fraction of individuals at risk $f_R$ and the fraction of individuals that are not cautious $f_{nC}$ in the population. For each pair $\left( f_R,f_{nC} \right)$, we checked two extreme couplings; (i) minimal coupling, where the overlap (denoted by $\Phi$) between the group of individuals at risk, and the group of individuals that are taking risks is minimal, i.e., $\Phi=\Phi_{min}=max\left(f_R+f_{nC}-1,0 \right)$. (ii) maximal coupling, where the overlap between the group of individuals at risk, and the group of individuals that are taking risks is maximal, i.e., $\Phi=\Phi_{max}=min\left(f_R, f_{nC}\right)$.

To test this effect, in figure \ref{fig:HospCompare}
\subsection*{Caution influence on hospital capacity}
\begin{figure*}[h]
\centering
\includegraphics[width=1\linewidth]{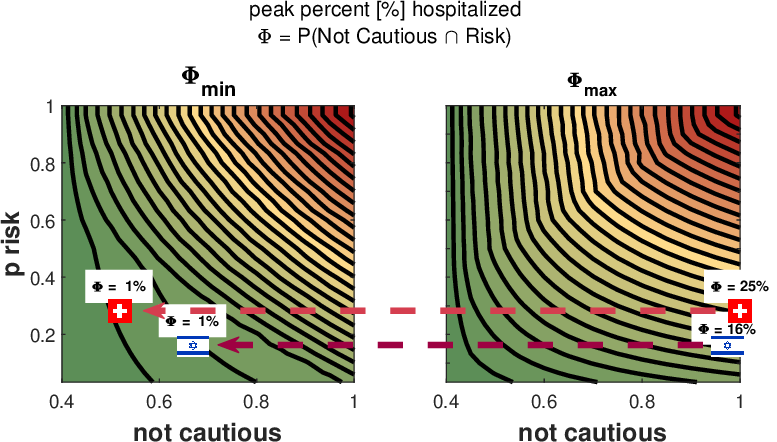}
\caption{
Hospital Stress for different levels of coupling of caution and risk. A Panel displays the peak hospital stress for a best-case scenario or maximal coupling. The maximal coupling occurs for minimal non-compliance, $ \Phi_{min} $. C panel displays the results in a worst-case scenario, of $Phi_{max}$, or maximal non-compliance. 
In the A, C panels the heat map, the browner tones denote higher hospitalization rates, greener tones denote lower stress. 
The largest difference between the two cases shown is at $P(risk) = 56 \%$, $P(Caution) = 25\%$, where the peak hospitalization changes from $2.4\%$ to $3.35\%$ by changing the $\Phi$ alone.
B panel denotes the change in caution and the change in $\Phi$ compliance, caused by vaccination, if we define vaccination as cautious behaviour. If all those at risk choose to vaccinate, the caution will increase, and the $\Phi$ will be $\Phi_{min}$.
The influence of the compliance $\Phi$, is clear from  comparing panels A and C. Note that for caution below $0.4$, there is no outbreak. By vaccinating and raising caution, Spain, Switzerland and Croatia moved below the epidemic threshold. Above each country is stated their $\Phi$ value, which is between that shown in Panel A and C. 
}
\label{fig:heatMapHosp}
\end{figure*}
In Fig. \ref{fig:heatMapHosp}, panel A and C we present the peak percentage of population hospitalized, for each pair of parameters namely the fraction of population at risk, and the fraction of population that is not cautious. Panel A is calculated for a minimal overlap scenario $\Phi=\Phi_{min}$. Panel C is calculated for the maximal overlap scenario $\Phi=\Phi_{max}$. In panel B we show the estimated level of risk and caution of selected countries. We show these values' change, from their vaccine policy, when we consider as proxy vaccination as cautious behaviour. This is consistent with the estimates shown in \cite{heterogenousMeansLessCovid}.

In Fig. \ref{fig:VaxSim} we also compared the estimated change in \(\Phi\), the overlap parameter and it's influence on the outbreak. We estimated vaccination data per age group in the population, and it's relative part within society for the risk and caution values, with vaccination being cautious behaviour using data from \cite{Mathieu2021}.

As seen from the figures, as more people at risk are more careful, the load on the hospital reduces. To exemplify the magnitude of this effect, consider a country with $50 \%$ of its population at risk, and $30 \%$ of its population not being cautious. If the size of the group of people at risk that are not caution is minimal ($\Phi=0$) then the peak hospitalization rate is less than $1 \%$. If the overlap is maximal, ($\Phi=30\%$), see equation \ref{eqn:Phi limits} the hospitalization rate increases by more than a factor of two, which is significant, since the level of service dramatically and non-linearly reduces upon breaching \cite{Eriksson2017}, \cite{French2021}.
\subsection*{Comparison to agent-based simulation}
\begin{figure}
\centering
\includegraphics[width=1\linewidth]{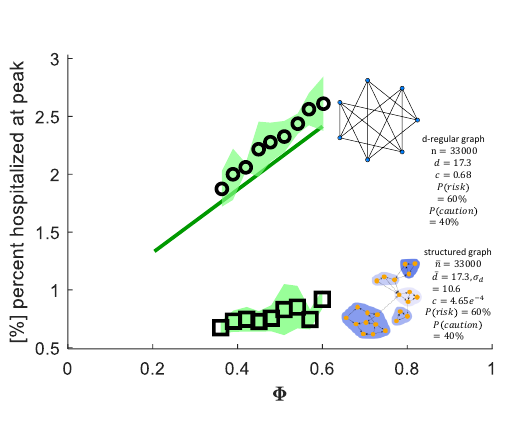}
\caption{
Peak hospitalization compared for different compliance levels \(\Phi\). The results of the ODE deterministic model are compared to results of stochastic models on two graph types. In the stochastic model, nodes on a graph simulated people in a society. The nodes are connected by a certain pattern. Initially, a few infected nodes gradually infect their neighbors, spreading the epidemic, simulating an outbreak. The types of graphs considered are a structured graph, reflecting household structures and d-regular graphs, where each node has a fixed number of neighbors. 
Here we show a case where the non-caution is equal to the part of the population at risk. The values of non-caution and risk shown here are equal \(60 \%\) and \(80 \%\) respectively. In the upper left panel we display where these results refer to in the heat maps shown in \ref{fig:heatMapHosp}. values shown display the parametric transition between the best-to-worst case scenario shown in \ref{fig:heatMapHosp}.
The solid lines are the ODE results, the values plotted over it are the stochastic models' results on a D-regular graph. For large enough graphs, the model based on d-regular graphs will converge to the ODE as the ODE offers a mean-field limit for the stochastic model.
The stochastic results are drawn from a sample of five simulations for each data point, with \(\sim 33000\) nodes or "agents". The Lower values denoted by squares are the peak hospitalization values for a structured graph. The structure reflects the key transmission channels and has a higher clustering coefficient than the D-regular graph, which lowers the outbreak speed by a factor of \(\sim 2\).
If the \(P(not cautious) = 0.4\) and \(P(Risk) = 0.2\), we find that in a d-regular graph there is an outbreak, with a \(0.8\%\) growth in infected numbers per day on average. However, in the structured graph there is an epidemic decline of \(0.9\%\) decrease in infected figures per day, irrespective of \(\Phi\). Here the epidemic was unable to spread in the structured graph, due to its larger clustering coefficient.
}
\label{fig:HospCompare}
\end{figure}
To further consolidate the ODE model, in Fig. \ref{fig:HospCompare} we also tested stochastic dynamics, using two types of social graphs. Random $d$-regular graphs, where each individual encounter exactly $d$ others at random, and a structured graph. These graphs' structure is chosen to match social connections of the state of Israel to simulated key infection channels. These connections consider distributions of: family size, school class size, number of co-workers at work, as well as random encounters. We found that the stochastic simulation on a d-regular graph, yielded similar results to the ODE simulation with the same parameters. The structured graph yielded a reduced hospitalization rate, an effect we attribute to its large clustering coefficient \cite{CM}. 
We find that the peak hospitalization rates are linear with the \(\Phi\) value in both models.
 
In \ref{fig:HospCompare} we tested the validity of our results with an agent-based simulation, where each person is an "agent", simulating the variance of an outbreak within a small group.
Each new node is connected to a fixed set of neighboring nodes, which becomes infected by one of them, and starts infecting others. 
This will simulate the ODE process, but this adds a level of randomness, as the ODE displays the average-case, whereas a slightly different result will appear each time with agents.
These agents are represented by nodes on a graph, connected by links. They are connected to either a fixed number of nodes (a "d-regular graph") or a graph linked in a distinct structure reflecting a society.
By imposing a structure on the graph, the structure itself also lowers the peak hospital stress, but the risk-caution coupling holds and still lowers peak hospitalization rates.
The results are noisy in an agent-based simulation, caused by variations in the outbreak due to stochastic simulation behaviour, whilst the numerical model only offers the average case scenario. 
Simulating this effect can help reflect and estimate variations in outbreaks caused by random noise. 
Here the d-regular graph's results converge to the ODE's predicted results when the number of nodes \(N\) is in the order of \(10,000\). 

\subsection*{Correlation of behaviour and perceived risk modify the number of hospitalized patients in response to Lockdown}

So far we discussed the effect of a fixed level of coupling between risk and cautious behavior. However, in reality, the level of risk aversion behaviour taken by an individual is dynamic and can change in time due to new information or a change of public opinion. To demonstrate the effect of a dynamic change in behaviour, motivated by a true event that occurred in March 2020. 

When a lockdown was forced for the first time in Israel, there was a naive expectation that the number of detected infections, and the number of hospitalizations, will show, after a delay, a marked change in slope. This prediction was based on simulating the effect of a lockdown by an abrupt discontinuous change in the infectivity parameter $\beta$. Such an effect did not occur, and its absence incited a mini-debate among analysts. 

Our explanation was based on the observation that the news reports coming from Italy and China, caused individuals at risk to avoid social contact, even before the lockdown was enforced slowing epidemic spread, see \cite{CC} \cite{chinaCovidArticle}, \cite{BFM}, \cite{NZ}. This made the effect of the lockdown appear "smeared" out and more gradual compared to the original expectation for an abrupt change in the slope. This is demonstrated in our model by modifying the caution factor to be a monotonic saturable function, e.g., a Hill function, of the number of hospitalized individuals in a state with a preceding outbreak.   
We were interested in testing the combined effect of a lockdown with elective switching towards cautious behaviour. Motivation for this scenario came from first lockdown imposed in Israel. Several analysts, including ourselves, where expecting the lockdown effect on hospitalization to appear  as a clear change of slope, which is was expected to be smoothed by natural stochastic delays in the time from infection to subsequent hospitalization. The data however showed no signs of such "knee", posing a riddle which we believe might be explained by pre-lockdown risk aversion behaviour of the fraction in the population that was at risk [GobermanPugatch].

\begin{figure}
\centering
\includegraphics[width=1\linewidth]{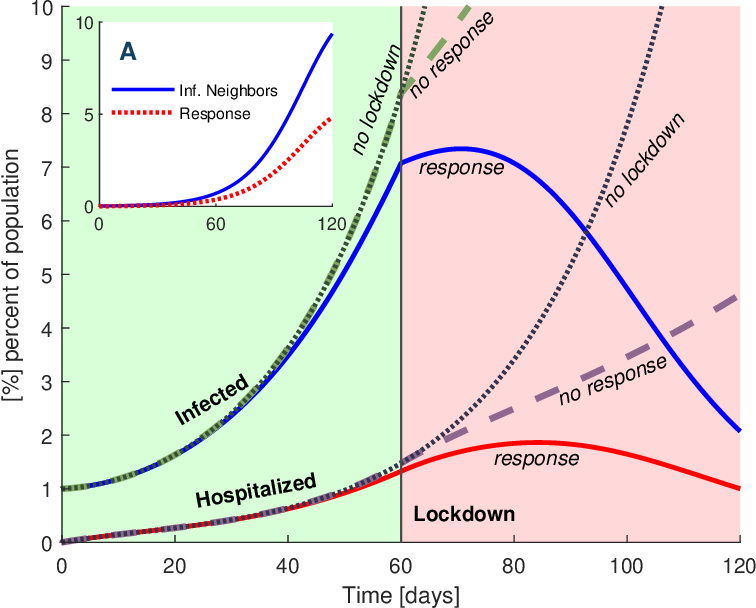}
\caption{
The number of people infected as a function of time with coupling to a neighboring country and Lockdown imposed after 60 days.
We compare the number of people infected and hospitalized, with or without elective response within the population, comparing it's influence on the efficiency of a lockdown in place.
In the A panel we see the outbreak's progression in a neighboring country, in a blue solid line. The local response to the outbreak, causing a transition towards cautious behaviour is marked in dotted red. So by \(T=120\) days into the outbreak, \(\sim5\%\) of the population has transitioned from non-caution to cautious behaviour. 
In the main panel, we see the influence of the transition towards caution, when combined with a lockdown. The x-axis denotes the time since the initial outbreak and the first case imported and the y-axis denotes the percent of people infected and hospitalized out of a population of 9 million. The dotted vertical line denotes a lockdown, put in place at \(T=60\) days.
Full lines denote the results of elective caution - people choose to transfer into the cautious group as a function of the number of cases in a neighboring country.
Blue full line marks the number of infected people, the Red full line marks the number of people hospitalized, when combined with the effect of a lockdown. We notice the lockdown causes a kink in the outbreak spread in the number of infected people. The influence of the lockdown on hospitalizations is more gradual. The combined effect of the lockdown and the transition towards caution, causes epidemic decrease.
Interrupted lines mark the local 'persistent' behaviour - where each group remains in the same position throughout the simulation. Green interrupted line denotes the number of people infected, the Gray interrupted line denotes the number of people hospitalized. The lockdown in this case fails to stop the outbreak, but is does manage to slows it significantly. The dotted lines mark the case of no lockdown or transition, where the outbreak grows unhindered.
}
\label{fig:Lockdown}
\end{figure}

In figure \ref{fig:Lockdown}, we plot the dynamics of an outbreak, assuming the local population take preemptive measures towards caution, hearing the news from another country, where the outbreak started a few weeks earlier.

To model this dynamic change in the coupling, we assumed the fraction of the population at risk starts to leak towards cautious behavior, prior to the lockdown, due to the news coming from countries that are ahead in the outbreak. 
We find that the coupling significantly lowers both the severity of the outbreak and it’s rate, and also causing the rate of change to be smoothed out, compared to the country that has a preceding outbreak.

In figure \ref{fig:Lockdown} we check the influence of 'rumor spreading', on outbreak rates, when combined with a lockdown.
We compare three cases---When a fixed number of people are cautious, compared to people responding to an outbreak in a neighboring country and choosing to respond to perceived risk, with neither effects implemented as a reference. We combine this with a lockdown put in place sixty days after the first cases emerge.
\subsection*{Vaccination can lead to reduced perceived risk and hence to an increase in risk-taking behaviour}
\begin{figure}
\centering
\includegraphics[width=1\linewidth]{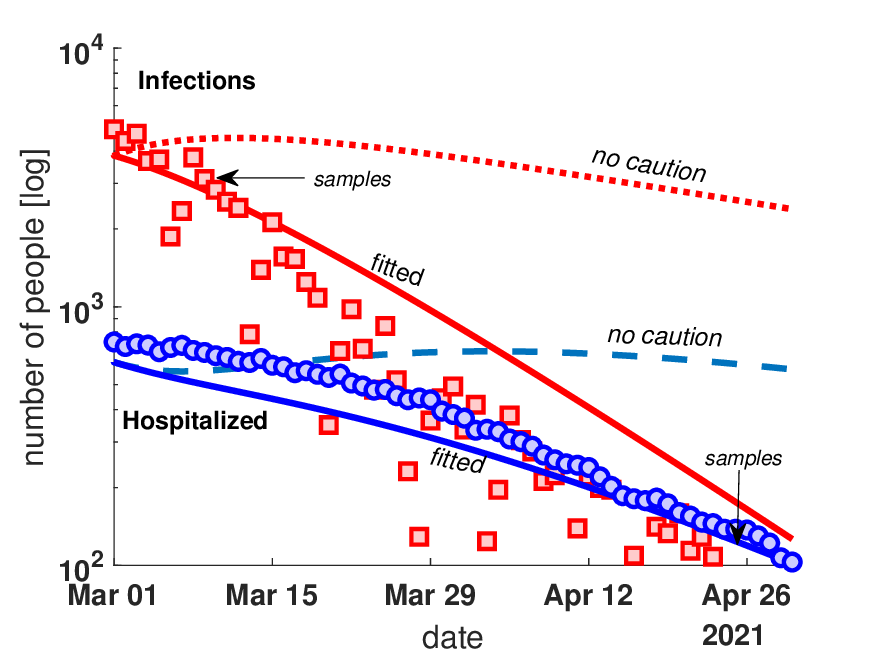}
\caption{
Fitting our model to real-world vaccination data. We fit our model to infection and hospitalization data collected from Israel in March-April 2021. We equate vaccination to cautious behaviour in our model. We compare the fit to a relapse in caution, which we theorize occurs due to a heightened sense of security after vaccination. We assume that after vaccination, people will be less willing to wear masks, thus reducing their caution.
In the figure there are two clusters of data. Infection data, above, denoting the number of daily new cases. Below, hospitalization rates, denoting the number of people hospitalized with Sars-nCov-2.
In each cluster, we see the samples, marked in data points, squares---infections, circles---hospitalization rates for real-world data.
The solid lines denote the results when fitted to our model. The dotted lines denote degraded behaviour, when upon vaccination, \(10\%\) of those vaccinated turn towards less cautious behaviour. The lower caution is expressed in doubling the chance of infection.
The Y-axis is a logarithmic scale of the number of people.
The X-axis is the date, starting on march 1st 2021.
}
\label{fig:VaxSim}
\end{figure}
To further demonstrate the effect of coupling of actual risk and behaviour we studied the effect of vaccination on behaviour. We wanted to test the effect of lax behaviour after being vaccinated, due to a higher sense of security. It can reduce the effect of the vaccine \cite{CovidInfectionOverTime}. 
We fitted our model to data from Israel, between March-21 the end of April, 21, i.e., the decline of the 3rd wave. We then modulated the infectivity parameter, to study the effect of remiss in behaviour towards more risk, once vaccinated. The effect can partially negate the vaccine. It is linear in the decay rate - if the vaccine quarters the chance of infection and risky behaviour doubles it - the vaccine will only halve infectivity at best. The more efficient the vaccine, the less sensitive the decline rate to a change in behaviour. Nevertheless, we see that in the case studied, a $10\%$ reduction in caution can lead to a $50\%$ percent slower decline rate (see Fig, X) \ref{fig:VaxSim}. 

Next, we studied the effect of becoming less cautious after vaccination. We define the risk group to be those not-vaccinated, and those vaccinated are at lower risk.
The vaccinated, non-risk group, may have a subsequent drop in caution. Non-vaccinated individuals, at higher risk now, may have an increase in caution when compared to those at lower risk.

In \ref{fig:VaxSim} we initially fit real-world data from Israel, artificially assuming all infected cases are detected. 
We then parametrically decreased the matching of caution and risk by \(30\) percent.
\(30\) percent of those at risk transferred from caution to non-caution.
\(30\) percent of those not at risk transferred to caution from non-caution.
Then we find that while the vaccination lowers the infection by \(82\%\) in our model, the matching of lower caution with those at actual higher risk causes there to be a new outbreak.

Another example of the effect of coupling can be measured in the emergence of a virus in a partially vaccinated population. We measure the outbreak for various vaccine efficiency to test the influence of caution on those not vaccinated. This shows that if part of the population is vaccinated, the outbreak can slows significantly if those who aren't vaccinated react by better protecting themselves. 

\section*{Methods}
\begin{figure}
\centering
\includegraphics[width=1\linewidth]{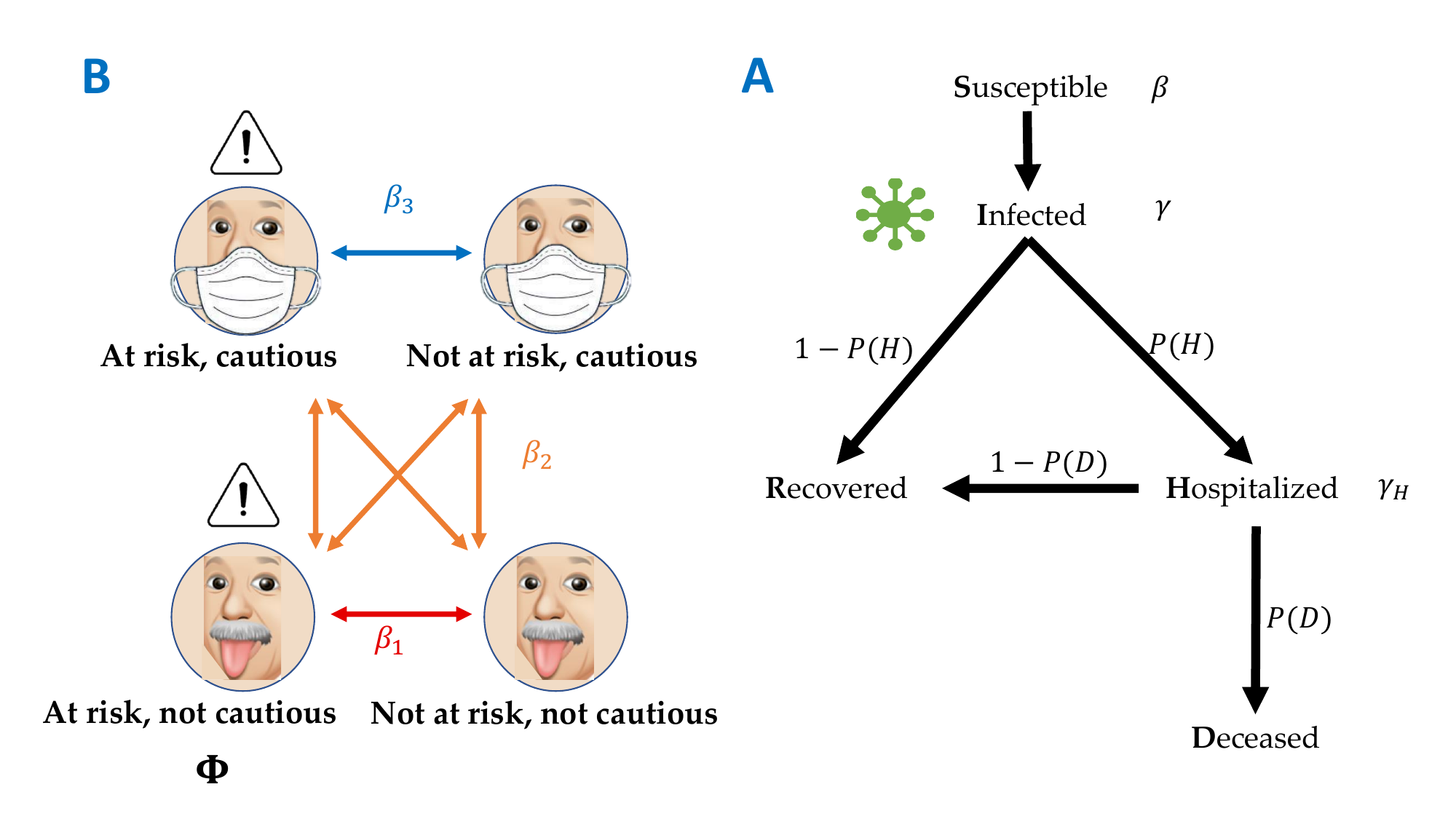}
\caption{
\textbf{States in our Model}
\\\hspace{\textwidth}
Panel A: We divide the population into four quadrants, split by two criteria: whether one chooses cautious behaviour or not and whether or not one is at high risk of severe disease once infected.
Each person stays in the group they started in throughout the entire simulation, except in the version shown in \ref{fig:Lockdown}.
The mask symbolizes the people who are cautious, the warning triangle denotes higher risk when infected.
Upper left: High risk group, and cautious. upper right: High risk, not cautious.
Lower left: Low risk group, and cautious. lower right: Low risk, not cautious.
We denote the parameter \(Phi\) as the portion of the group at risk and not cautious.
\( \beta_{1,2,3} \) denote the differing infection rates according to caution: \( \beta_1 = 1\) denotes infection rates if both people meeting are at not cautious, \( \beta_2 = 0.85\) is if only one person in the meeting is cautious and \( \beta_3 = 0.05\) if both are.
\\\hspace{\textwidth}
Panel B: The five dynamic states people transition through during the epidemic: 
1. Initially, all are susceptible to the epidemic. The time required to move to infected is exponential with parameter \(\beta\). 
2. Infected (no latent period assumed). When infected, one is also infecting others coming into contact, with each contact having different infection rate caused by varying exposure (caution). From here one proceeds either to hospitalization or to recovery. The duration spent here is drawn from exponential distribution with rate \(\gamma\).
From infectious, when time here has elapsed transition to hospital occurs with probability \(P_H\). \(P_H\) differs is if the person is at risk or not. otherwise to recovered group, with probability \(1-P_H\). 
3. Hospital, straining critical health resources, from here one can move to recovery or deceased state. Time spent here is drawn from exponential distribution with rate \(\gamma_H\), which is different for those at risk. Upon transition out of hospital, transfer to deceased group with probability \(P(D)\), which is dependant on risk, otherwise to recovered group with probability \(1 - P(D)\).
4. Recovered group, in the duration of the simulation we don't consider re-infection, so people here stay in this final state.
5. Deceased.
\\\hspace{\textwidth}
The Risk groups have \(P(H_R)\) and \(P(D_R)\) significantly higher than that of non-risk groups.
\\\hspace{\textwidth}
Overall there are 5 stages and 4 groups so the population state vector is of size \(4 * 5 = 20\).
}
\label{fig:ModelStates}
\end{figure}

We model a population with a given portion of it at risk, for explanation see Fig. \ref{fig:ModelStates}. Risk is modeled as an increased probability of hospitalization and death for an infected individual. To evaluate the effect of coupling between risk and caution we parametrically scan the fraction of population that is both at risk and does not behave cautiously--- $\Phi$. 

For each portion of population at risk, and fraction not cautious, we use a set of coupled differential equations \ref{SEquationFull} to model the number of infections, hospitalizations and deaths.

In figure \ref{fig:ModelStates} we describe our model.  We use an SIR-based ODE model, which classically comprises of susceptible-infected-recovered. We add the states of hospital and deceased, so we have an "SIRHD" model. 
Each person starts as S (susceptible), then upon infection transfers to I (infected). From here either R (recovered) or H (hospitalized). From H one can proceed to R (recovered) or D (deceased). The D and R states are final, i.e there is no second infection. We also assume there is no death without hospitalization first.
\subsection{Equations defining the model}
We list the equations for the susceptible group at full length, the rest continue in a similar fashion, listed in full in SI . 
From the susceptible state, people get infected and transfer to I group at a rate proportional to \(S \times I\). The risk cautious groups become infected with \(\beta_{1, 2}\), the not-cautious groups are infected at a rate of \(beta_{2, 3}\) if the infector is cautious or not cautious respectively.
\begin {equation}
\resizebox{0.95\hsize}{!}{ $ \frac{dS_{RC,nRC } }{dt} =  S_{RC,nRC}\overbrace{((I_{RC} + I_{nRC}) \cdot \beta_{CC}}^{infections\:from\:cautious} + 
\overbrace{ (I_{SnC} + I_{nSnC}) \cdot \beta_{CnC})}^{infections \:from\: non-cautious} $ }
\end {equation}
\begin{equation}
 \resizebox{0.95 \hsize}{!}{ $ \frac{dS_{SnC,nSnC } }{dt} =  S_{SnC,nSnC }((I_{RC} + I_{nRC}) \cdot \beta_{CnC} -  (I_{SnC} + I_{nSnC}) \cdot \beta_{nCnC} ) $ }
\label{SEquationFull}
\end{equation}
In I group which infects the S and transition to R or H at a rate proportional to the size of I.
Continuing in a similar fashion, we will display values all four states: \[S,nS,C,nC\] as a vector, and shorten the display:
\begin{equation}
\frac{d\vec{I} }{dt} = +\overbrace{\beta \cdot \frac{d\vec{S}}{dt}\cdot \vec{I}}^{new \:infections} - \overbrace{\gamma \cdot \vec{I}}^{recovery} \\
\end {equation}
The H group receives people from I and proceed to either R or D.
\begin{equation}
\frac{d\vec{H}}{dt} = +\overbrace{ \gamma \cdot P(H_{S,nS}) \cdot \vec{I}}^{new \:hospitalized} - \overbrace{\gamma_{H} \cdot \vec{H}}^{leaving \:hospital}\\
\end {equation}
The R group receives recovered people from H and I.
\begin{equation}
\frac{d\vec{R}}{dt} = +\overbrace{ \gamma \cdot (1-P(H_{S,nS})) \cdot \vec{I}}^{new \:recovered} +\overbrace{\gamma_{H} \cdot (1-P(D_{S,nS}) \cdot \vec{H}}^{recovered \:from \:hospital} \\
\end {equation}
The D group receives people from H.
\begin{equation}
\frac{d\vec{D}}{dt} =+\overbrace{\gamma_{H} \cdot (P(D_{S,nS}) \cdot \vec{H}}^{deceased \:from \:hospital} 
\end {equation}
In this model, at the end of the simulation, the population is split between Susceptible, Recovered and Deceased.
In a well-mixed model, with low clustering, this is mostly dependent on the growth rate, defined in equation \ref{eqn:GrowthEqn}.

\subsection{metrics for our model}
As \(R_0\) is difficult to analyse consistently comparatively between different models, 
we use the parameter \(G\) growth rate for analysis.
We use the equation:
\begin{equation}
 G = \beta/\gamma
\label{eqn:GrowthEqn}
 \end{equation}
Note that the epidemic threshold \(R_0 > 1\) being an outbreak, \(R_0 <1 \) denoting epidemic decline, 
can still be applied to parameter \(G\), the threshold becomes \(G > 0\) and \(G < 0\) respectively.

\subsection*{Compliance parameter \(\Phi\)}
The parameter we use to describe compliance we define as:
\begin{equation}
\Phi = P \left( not cautious \cap risk \right) 
\label{eqn:Phi eqn}
\end{equation}
\(\Phi\) is limited to be between the following two values:
\begin{equation}
\max \left( P(risk)+P(caution), 0\right) \leq \Phi \leq \min \left(P(risk), P(cautious) \right)
\label{eqn:Phi limits}
\end{equation}

\subsection*{Parameters of the model} 
At each stage, the time one will spend in a state is generated exponentially. Here we define the rates of transition from each state, based on estimates from \cite{XE}.
\(\beta\): the infection rate, changes from a susceptible person meeting an infectious person, with 3 possible options: cautious-cautious, both the infector and infcetee are cautious, cautious-not cautious, infector isn't cautious, infectee is (or vice versa, they are equal in this case), and not cautious - not cautious.
\(\beta_0 \) denotes the rate of new infections, ignoring the changes caution causes. Needs to be determined relative to \(\gamma\), as showed in \ref{eqn:GrowthEqn}, to give \(G = \beta_0/\gamma\).

We use: \([\beta_1, \beta_2, \beta_3] = [ I_CC, I_CnC, I_nCnC] = [0.05, 0.15, 1 ] \) this estimate was chosen as it was assumed that by choosing to wear masks, i.e act cautiously, the infectious ability is lowered by 95 percent if both people meeting are wearing masks as shown in \cite{JH}.
We get the following equation:
\[ \beta' = \beta_0 \cdot (I_CC \cdot P(CC) + I_CnC \cdot P(CnC) + I_nCnC \cdot P(nCnC)) \]
With which we calculate the eventual effective \( \beta' \). 
 In the numerical model, the parameter is directly multiplied by the basic \(\beta_0\). The value of \(\beta_0\) is \(\frac{3}{10}\). In the agent-based model it is expressed through changing weights of edges in the graph.
\(\gamma\): the rate at which a person transitions from being infectious to later states of either hospital or recovery. The value of \(\gamma\) is \(\frac{1}{10}\). In this way, \(G\) is equal to \(\beta_0/\gamma= 3\).
\(N\) population size: the number of people in the simulation. It's value is 9 million to reflect the population of Israel. In the agent-based simulations, \(N = \sim 33000\), as it balances the population size and mean-case convergence with computational resources.
The part of the population at high risk in case of infection is a parameter marked \(P(R) \).
The probability of hospitalization is marked \(P(H)\). The value for risk group is \(P(H_{Risk} = 0.2\) and \(P(H_{not Risk}) = 0.02\) for individuals not at high risk. Whether or not one is hospitalized is determined for each individual randomly upon completion of time spent in infected state, which is occurs with rate paramter \(\gamma\). 
All the parameters are based on estimates from \cite{XE} for Sars-Cov-2 and we assume no latent period as it doesn't change the outcome for an outbreak, only incurs a delay in our model. We don't consider symptoms, or symptomatic changes in behaviour.
The time spent hospitalized is an exponential rate parameter marked \(gamma_H\). The value of \(gamma_H\) is \(\frac{1}{20}\) for risk and non-risk groups. 
The probability of death upon hospitalization, value for risk group is \(P(D_{Risk}) = 0.2\) and not at high risk is \(P(D_{not Risk}) = \frac{0.2}{5}\).
The initial number of infected agents is \(1\%\) to ensure an initial outbreak, in the ODE it is \(\frac{1}{1000000}\) as it is numerically stable. 

\subsection*{Agent-based simulations}
We next run our model on an agent-based simulation, to see how local random behaviour differs to the macro behaviour observed by the ODE solvers.
In this model, each node in a graph is an "agent", that thinks for itself, and views connections via links in a network on a mathematical graph.
The underlying graph structure changes the connections each person or node has, and will change the way an epidemic will spread on a graph, as shown in \cite{ME}, \cite{ML}, \cite{MN}, \cite{HouseHoldTransmissionOfCovid}.
We choose two graphs to compare: First, a d-redular graph, where each person has a fixed number of neighbors randomly selected within the population. When running a simulation on this type of graph, the outbreak will converge (given enough nodes and large enough D) to an average behaviour like that of the ODE. 
We also build a graph that reflects a society's structure, and it's key factors in spreading epidemics. The strength of a connection changes depending on how closely they meet, and changes the infection rate, \cite{AG}. 
We describe interactions as three classes: Home interactions (strong), school or Work (medium), or random (weak).
Each person is either young and is at lower risk, or adult and may be at higher risk. 
To develop the graph we use use the method: 
Step 1: Type "a" - build a household or "family" with strong connections, selected to behave like Israeli households. 
Step 2: Type "b" - connect the adults to workplace colleagues and children to school peers.
Step 3: Type "c"- add random connections considering other random interactions from service providers, transport, etc. 
For a full description of the household size distributions please see SI [x].
The weights are $[w_a=2.5$, $w_b=0.7$, and $w_c=0.4$. The mean and variance of the number of edges is \(\bar{d}=17.3, \bar{Var(d)} = 10.5\) respectively.
For example: node 112 and node 113 share a household, therefore their connection is of type "a" and that weight is \(w_a = 2.5\).
To build the d-regular graph, all edges are weighted equally. The graph's weights are equal to the mean weight in the comparative structured graph. The d---or number of edges---the d-regular has, is equal to the mean number of edges in the structured graph, with a value in our case of \(d=\sim17.3\). 
We then compare the outbreaks in each graph type to the ODE solver we used earlier, in \ref{fig:HospCompare}, and run 5 times. For details see SI. 
In figure \ref{fig:Lockdown}, we show the combined effect of preemptive caution, combined with a lockdown enforced. The coupling to the neighboring country is \(0.9\). For example, if \(10\%\) of the neighbors are infected, \(0.9 \cdot 0.1 = 0.09\), the transition rate of those not cautious transfer towards cautious behaviour over time would be of \(9\%\). The parameters of the neighboring country are identical, however no lockdown or mitigation efforts are made.
\subsection{Fitted data to our model}
The data we used in \ref{fig:VaxSim} was taken from [X].
We fitted our model's parameters of vaccine efficacy and \(P(Caution)\) using the estimated data. The model we used was least-squares on these two parameters, with the \(L^2\) error norm.
\section*{Discussion}
The effect of coupling between actual risk and risk aversion behaviour, or lack thereof, affects the peak hospitalization and also the mortality toll of an epidemic. These effects can be estimated both statically---assuming a fixed overlap parameter $\Phi$---or dynamically---assuming $\Phi$ to be state dependent or to depend explicitly on time. Our appriach allows for estimation that is strictly bounded from above and from below. This enables qualitative understanding of the importance of this effect, and thus also on the need to take it into account.

State dependent $\Phi$ is also of interest. Since risk aversion behaviour is a result of risk awareness, there an interesting interplay between risk awareness or alternatively, misguided risk awareness, which can be called epistemological risk, and risk itself, which can be called ontological risk. This can lead to very rich and complex dynamics of epidemics, dynamics that couple rumor spreading with disease spreading. It is thus of interest to write down such models using e.g. voter models or rumor spreading models on networks, and in parallel to use a network of social contacts---not necessarily the same network---in which infectious contact strength are modulated according to the belief state of the infectious agents. This literal belief-propagation coupled to infection propagation can also be a function of the infection-fatality rate (IFR). Apparently, in infectious diseases with an extremely large IFR such as Ebola, rumors might cause a dramatic reduction in social contacts. It seems than in infectious diseases with a small IFR, there is more room for a reverse effect, to the extent that spread of disbelief might encourage people to even ignore risk.
\showmatmethods{} 

\acknow{I thank Dr. Rami Pugatch for his dedicated supervision, guidance, and generous sharing of knowledge throughout the completion of this work.}

\showacknow{} 

\bibliography{citations.bib}
\end{document}